\documentclass[conference]{IEEEtran}
\IEEEoverridecommandlockouts
\usepackage{cite}
\usepackage{amsmath,amssymb,amsfonts}
\usepackage{algorithmic}
\usepackage{graphicx}
\usepackage{textcomp}
\usepackage{xcolor}
\usepackage{mathtools}
\usepackage[caption=false]{subfig}
\usepackage{url}
\usepackage{soul}
\usepackage{rotating}
\usepackage{array}
\usepackage{multirow}

\IEEEpubid{\begin{minipage}{\textwidth}\ \\[12pt]
978-1-6654-3886-5/21/\$31.00 \copyright2021 IEEE
\end{minipage}}

\def\BibTeX{{\rm B\kern-.05em{\sc i\kern-.025em b}\kern-.08em
    T\kern-.1667em\lower.7ex\hbox{E}\kern-.125emX}}

\title{Towards General Models of Player Experience: \\A Study Within Genres
\thanks{This project has received funding from the EU’s Horizon 2020 programme under grant agreement No 951911, and from the University of Malta internal research grants programme Research Excellence Fund under grant agreement No 202003.}
}

\author{\IEEEauthorblockN{David Melhart}
\IEEEauthorblockA{\textit{Institute of Digital Games} \\
\textit{University of Malta}\\
Msida, Malta \\
david.melhart@um.edu.mt}
\and
\IEEEauthorblockN{Antonios Liapis}
\IEEEauthorblockA{\textit{Institute of Digital Games} \\
\textit{University of Malta}\\
Msida, Malta \\
antonios.liapis@um.edu.mt}
\and
\IEEEauthorblockN{Georgios N. Yannakakis}
\IEEEauthorblockA{\textit{Institute of Digital Games} \\
\textit{University of Malta}\\
Msida, Malta \\
georgios.yannakakis@um.edu.mt}
}

\begin{document}

\maketitle

\begin{abstract}
%Can abstract gameplay metrics realise general models of experience on par with game-specific ones? 
To which degree can abstract gameplay metrics capture the player experience in a general fashion within a game genre? In this comprehensive study we address this question across three different videogame genres: racing, shooter, and platformer games. Using high-level gameplay features that feed preference learning models we are able to predict arousal accurately across different games of the same genre in a large-scale dataset of over $1,000$ arousal-annotated play sessions. Our genre models predict changes in arousal with up to $74\%$ accuracy on average across all genres and $86\%$ in the best cases. We also examine the feature importance during the modelling process and find that time-related features largely contribute to the performance of both game and genre models. The prominence of these game-agnostic features show the importance of the temporal dynamics of the play experience in modelling, but also highlight some of the challenges for the future of general affect modelling in games and beyond.

% GENRE VERSUS GAME: GENERAL FEATURES SUCH AS TIME 
\end{abstract}


\section{Introduction}

% \todo[color=green]{AL reviewed, good job David}

Artificial general intelligence and artificial psychology define two critical long-term goals of artificial intelligence (AI). The intersection of the two would enable artificial systems to perform affect-based interactions in general settings. While games (board or digital) define the dominant application area for the study of general AI, limited emphasis has been given to the ways general AI systems are possible in games beyond the task of gameplaying \cite{perez2016general,togelius2016general}, including systems that create or even model player experience in a general fashion \cite{yannakakis2018artificial}. Arguably, studying \textit{general models of player experience}---which aim at predicting the experience of play in a game-independent way---is still in its infancy. The handful of examples in this vein are limited by ad-hoc game testbeds, and experience models that are built on small-scale corpora \cite{shaker2015towards,shaker2016transfer,camilleri2017towards}.

Motivated by the lack of a comprehensive study on general player experience modelling, this paper explores the degree to which player experience can be modelled across games of the same genre in a general fashion. We assume that there exist features of play that are able to transfer aspects of player experience across games of the same genre. We also assume that such features can be used to build accurate models of player experience in a general fashion within a genre. 

%David: you need to refer to this figure from somewhere in the text/ Also: what is pairwise transformation? Can you describe the pipeline in the caption please? The figure is nice but largely unexplained
\begin{figure}[!tb]
\centering
\includegraphics[width=1\linewidth]{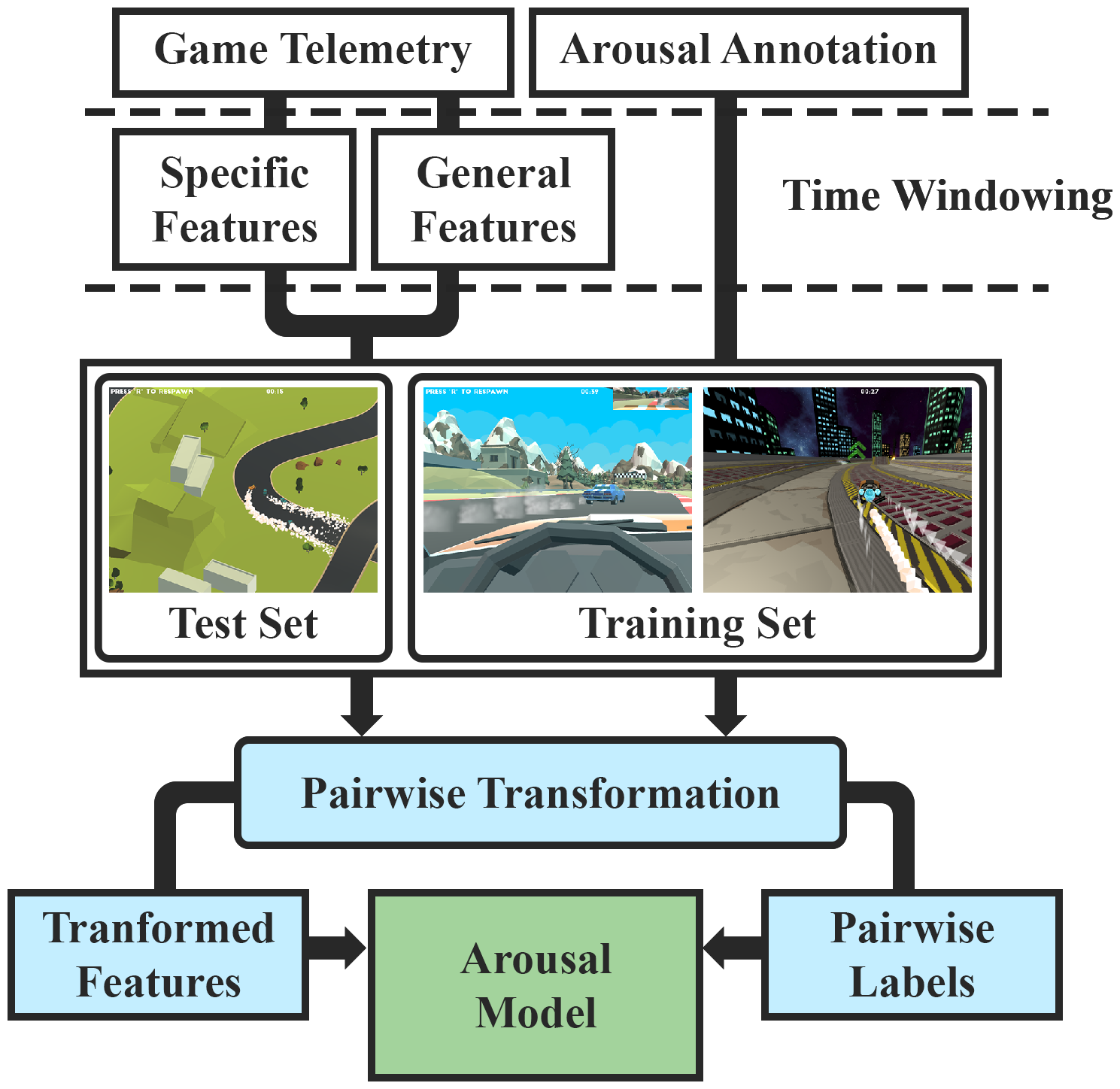}
\caption{Genre-based modelling pipeline for modelling arousal across different games of the same genre as presented in this study. Both genre-specific and general features are extracted from the raw telemetry. Models are trained on data from two games and tested on an unseen game within the genre. Preference learning is applied by using a Pairwise Transformation, in which the ranking problem is reformulated as binary classification of pairwise labels (see Section~\ref{sec:methods:pl} for more details).}
\label{fig:pipeline}
\end{figure}

To test our hypothesis, we first design features \emph{specific} to each genre that can predict player experience within a game with high accuracy. Then we examine whether certain ad-hoc designed features that contain general information about gameplay can act as reliable \emph{general} predictors of player experience across games of the same genre. We use the \emph{Affect Game AnnotatIoN} (AGAIN) dataset \cite{melhart2021affect}, which includes telemetry and annotations of arousal for almost 1,000 play sessions of nine different games across three different genres (\textit{Racing}, \textit{Shooter}, and \textit{Platformer}). We employ random forests for preference learning in order to build models that predict arousal both in a specific game and across unseen games of each genre (see Figure~\ref{fig:pipeline}). Viewing player arousal modelling as a relational learning problem \cite{yannakakis2018ordinal}, we test the capacity of the models to predict the \textit{change in arousal} in a short time-frame compared to the previous session history. The key results show that the ad-hoc designed general features manage to predict the change of arousal with up to $74\%$ accuracy on average across all genres and $86\%$ in the best cases within shooter games. The core findings of the paper suggest that we can design general features that can predict player experience across unseen games of the same genre with high accuracy. More importantly, such features perform equally well compared to features that are tailored to predict player arousal within the same game that they were trained on. 

To the best of our knowledge, we examine genre-based general modelling of player arousal through game telemetry for the first time. Only one similar study examined general arousal modelling \cite{camilleri2017towards}. However, here we take a more systematic approach and examine our results in the context of three different videogame genres: racing, shooter, and platformer games. We also reexamine ad-hoc gameplay features through an analysis of feature importance, explaining the results and process of our machine learning models. Our results highlight the importance of a similar temporal dynamic between games. This revelation puts previous general affect modelling approaches into context and foreshadows future challenges in the field.

\section{Background}\label{sec:background}

% \todo[color=green]{AL reviewed}
This section highlights related work on modelling players' affective states (Section \ref{sec:background:modelling}), and our {ordinal} approach to emotion modelling (Section \ref{sec:background:ordinal}).

\subsection{Player Affect Modelling}\label{sec:background:modelling}

The field of games user research can generally be divided between static profiling and dynamic modelling \cite{yannakakis2018artificial}. While the former focuses mostly on high-level data aggregation \cite{drachen2009player} and pattern discovery \cite{makarovych2018like}, the latter involves predictive modelling. These modelling tasks can be further broken down into behavioural (i.e. {what the player does}) and affective (i.e. {how the player feels}) approaches. Examples for the former include behaviour \cite{bakkes2012player} and churn prediction \cite{viljanen2018playtime}, while examples for the latter include experience \cite{makantasis2019pixels,melhart2020moment}, motivation \cite{melhart2018towards}, and affect modelling \cite{lopes2017modelling,camilleri2017towards}. Since most of these studies rely on supervised machine learning, their main limitation is their data needs. Many studies focus on ad-hoc testbeds and game-dependent models. While the resulting models are useful for understanding how players interact with already published games, these models do not generalise well to unseen ones. 

To answer this issue, general affect modelling \cite{togelius2016general} aims to create pre-trained models which can be applied to unseen games. If successful, such models can reduce the data needs of new projects. While research has begun in this field, studies in the literature are still rather sparse. For instance, Shaker \textit{et al.} investigated manual \cite{shaker2015towards} and automated feature mapping \cite{shaker2016transfer} through the use of transfer learning. While transfer learning offers a robust approach, interestingly, other studies have been just as successful in applying domain knowledge to hand-craft high-level general features of gameplay. Camilleri \textit{et al.} used game-agnostic features such as {playtime} and encoded {valence} as {goal oriented} and {goal opposed} events to model arousal across games \cite{camilleri2017towards} with moderate success. Similarly, Bonometti \textit{et al.} used {activity count} and {diversity} to abstract gameplay and model general engagement across six games \cite{bonometti2020theory}. However, a general limitation of these studies is the ad-hoc set of testbed games, which are often limited in scope or fall too far from each other. 
In this paper, we take a more structured approach to general modelling and investigate the robustness of domain-specific general features created in a top-down manner. As opposed to previous studies, which used ad-hoc setups, we investigate the proposed approach in three different genres, over nine different games.

\subsection{Ordinal Player Modelling}\label{sec:background:ordinal}

Ordinal affect modelling aims to capture the relative processes behind emotional responses \cite{yannakakis2017ordinal,yannakakis2018ordinal}. Human cognition is prone to temporal biases \cite{damasio1994descartes} such as {anchoring} \cite{seymour2008anchors}, {habituation} \cite{solomon1974opponent}, {adaptation} \cite{helson1964adaptation}, and other {recency effects} \cite{erk2003emotional}. Therefore, focusing on the relative differences rather than absolute judgements can lead to more reliable observations and more robust predictions \cite{yannakakis2018ordinal}. In the field of games user research, several papers contribute to a growing body of research proving the effectiveness of this approach; see \cite{martinez2014don,yannakakis2015grounding,yannakakis2015ratings,lotfian2016practical,melhart2018study,yannakakis2018ordinal} among many. This approach evidently increases the inter-rater reliability and consistency of data annotations \cite{yannakakis2015grounding,yannakakis2015ratings}, and yields models that have a higher generality across affective corpora \cite{melhart2018study} and dissimilar videogames \cite{camilleri2017towards}.

A common issue with ordinal affect modelling is the lack of sufficiently labelled datasets. Because collecting pairwise comparisons through forced-choice surveys can be labour intensive (due to the number of comparisons growing quadratically when new options are introduced), most studies focus on traditional rating methods such as Likert scales. While absolute ratings can be converted to ordinal labels \cite{yannakakis2018ordinal}, bounded scales come with their own limitations \cite{yannakakis2015ratings}. A good compromise is to collect unbounded ratings, which can still be interpreted in an ordinal fashion but preserves the relative relationship between data points \cite{yannakakis2017ordinal}. 
%Developed originally by Lopes \textit{et al.}, {RankTrace} is an annotation tool aimed to capture temporal and relative relationships of affect \cite{lopes2017ranktrace}. Lopes \textit{et al.} suggest four different metrics to process continuous, ordinal annotation. These are the a) the mean, b) the {area under the curve}, c) the {amplitude}, and d) the average gradient of the annotation. While the mean and average gradient are more intuitive (average value and average relative acceleration), {area under the curve} and {amplitude} can be harder to contextualise for all features. Camilleri \textit{et al.} adopts a similar approach but reduces the number of metrics to two: an {absolute} (mean) and a {relative} (average gradient) \cite{camilleri2017towards} measure. While in their study, Camilleri \textit{et al.} compares model performance based on the processing of the output, here we take a different approach. We look at the two metrics as two distinct tasks. 
Inspired by the studies of Lopes \textit{et al.} \cite{lopes2017ranktrace} and Camilleri \textit{et al.} \cite{camilleri2017towards}, we collect arousal in an unbounded continuous fashion and via the {mean} value within a time window to predict changes in arousal.

\section{The AGAIN Dataset}\label{sec:games}

\begin{figure}[t]
\centering
\includegraphics[width=\linewidth]{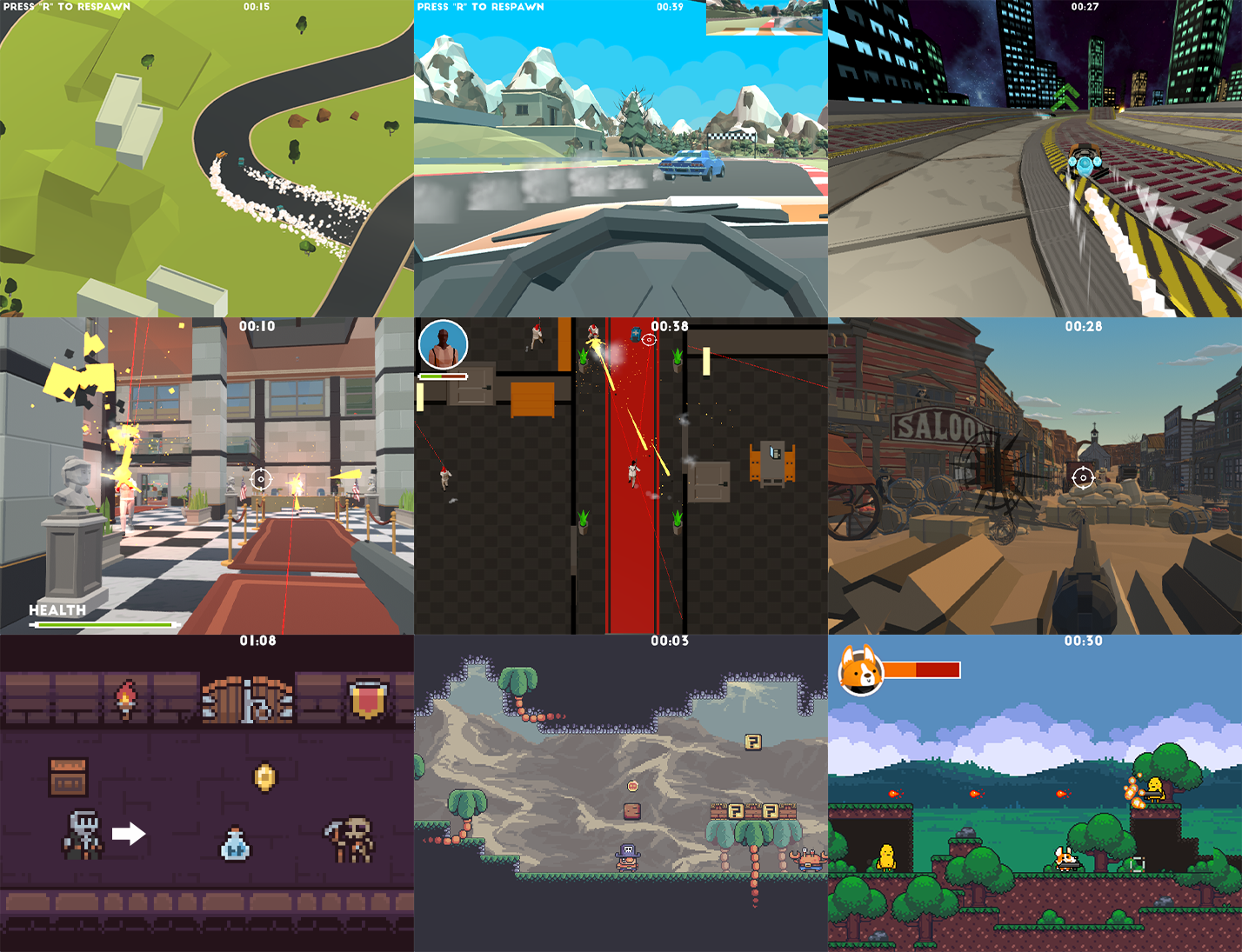}
\caption{The 9 games of the AGAIN dataset, one genre per row. Top row is \textit{racing} games, mid row is \textit{shooter} games and bottom row is \textit{platformer} games.}\label{fig:games}
\end{figure}

% \todo[color=green]{AL reviewed}

% \begin{figure*}[t]
% \centering
% \subfloat[TinyCars]{\includegraphics[width=0.3\textwidth]{img/games/TinyCars.png}\label{fig:game:tiny}} \quad
% \subfloat[Solid]{\includegraphics[width=0.3\textwidth]{img/games/Solid.png}\label{fig:game:solid}} \quad
% \subfloat[ApexSpeed]{\includegraphics[width=0.3\textwidth]{img/games/Apex.png}\label{fig:game:apex}} \\
% \subfloat[Heist!]{\includegraphics[width=0.3\textwidth]{img/games/Heist.png}\label{fig:game:fps}} \quad
% \subfloat[TopDown]{\includegraphics[width=0.3\textwidth]{img/games/TopDown.png}\label{fig:game:topdown}} \quad
% \subfloat[Shootout]{\includegraphics[width=0.3\textwidth]{img/games/Shootout.png}\label{fig:game:gallery}} \\
% \subfloat[Endless]{\includegraphics[width=0.3\textwidth]{img/games/Endless.png}\label{fig:game:endless}} \quad
% \subfloat[Pirates!]{\includegraphics[width=0.3\textwidth]{img/games/Pirates.png}\label{fig:game:platform}} \quad
% \subfloat[Run'N'Gun!]{\includegraphics[width=0.3\textwidth]{img/games/RunNGun.png}\label{fig:game:gun}}
% \caption{Start screens of the nine games included in the AGAIN dataset, showing the game's rules and players' controls.}\label{fig:games}
% \end{figure*}

This study employs the AGAIN dataset\footnote{The full dataset is available at \url{https://again.institutedigitalgames.com/}}, which was designed to provide a diverse and robust database for general affect modelling in the domain of videogames \cite{melhart2021affect}.
The raw dataset includes $1,116$ playthroughs; after cleaning and preprocessing, the clean dataset (used in this paper) includes 122 participants and 995 playthroughs. More information on the games, the cleaning process, and the dataset are found in \cite{melhart2021affect}.

\subsection{Games}

The AGAIN dataset includes 9 games in total; 3 games for each of the \textit{racing, shooter}, and \textit{platformer} genres (see Fig.~\ref{fig:games}). The games were designed as casual representations of popular contemporary and classic games. Because the games had to fit into a 2-minute playtime, they are simplified and resemble a mobile game experience rather than console or PC gameplay. Nevertheless, the games were designed to provide a more realistic testbed with more contemporary aesthetics.

\subsubsection{Racing} 

The AGAIN dataset includes three car-racing games, where players have to navigate in a closed-loop track until the timer runs out. 
In the order that they appear in Fig.~\ref{fig:games},
the three games in this set are: \textbf{TinyCars}, a retro top-down racing game; %(Fig. \ref{fig:game:tiny}); 
\textbf{Solid}, a rally game; %(Fig. \ref{fig:game:solid}); 
and \textbf{ApexSpeed}, an arcade-like speed racer. %(Fig. \ref{fig:game:apex}). 
All of these games feature three opponents who race against the player. The control scheme of the games is quite consistent; {ApexSpeed} stands out as the car moves automatically on a preset track with lane swapping mechanics.

\subsubsection{Shooter}

The set of shooters includes games where the goal is to eliminate opponents using projectile weapons. 
In the order that they appear in Fig.~\ref{fig:games},
the games in this set are \textbf{Heist!}, a first-person shooter with health regeneration mechanics; %(Fig. \ref{fig:game:fps}); 
\textbf{TopDown}, a retro top-down shooter with unlimited ammo and health pick-ups; %(Fig. \ref{fig:game:topdown});
and \textbf{Shootout}, an arcade shooter. 
%(Fig. \ref{fig:game:gallery}).
All of these games involve mouse-aim; however {Shootout} stands out as in this game, the player has no health and is not able to move; the game is only played for score.

\subsubsection{Platformer}

In the platformer games of AGAIN, players have to navigate in a 2D environment, eliminate or evade opponents, and solve light spatial puzzles. 
In the order that they appear in Fig.~\ref{fig:games},
the set includes \textbf{Endless}, an endless runner;
%(Fig. \ref{fig:game:endless}); 
\textbf{Pirates!}, a classic Mario-clone;
%(Fig. \ref{fig:game:platform}); 
and \textbf{Run'N'Gun}, a retro shoot-em up.
%(Fig. \ref{fig:game:gun}).
Platform games are the most diverse in the dataset, with {Endless} featuring automatic forward movement (like {ApexSpeed}) and {Run'N'Gun} featuring weapon aiming.

\subsection{Dataset}\label{sec:dataset}

The clean AGAIN dataset consists of 122 players playing 995 sessions of 2-minute games \cite{melhart2021affect}.
The gender distribution of the participants skews towards men. One participant identified as non-binary, 43 as female, and 78 as male. The average age of participants was 33, ranging from 19 to 55. Most respondents were from the USA (100 participants); other countries were Brazil (10 participants), Italy (3), Canada (2), India (2), Czech Republic (1), Germany (1), and Romania (1). Most of the participants (114) are self-described gamers (either hard-core or casual) and play videogames daily or weekly. Most participants had either a PC, gaming console, or both and played a wide variety of game genres, including shooters, platformers, and driving games.

While the dataset includes both video footage and telemetry data, we focus on the latter in this study.
The dataset contains genre-specific telemetry features, which are largely shared across games within the same genre. AGAIN provides $33$, $35$, and $42$ genre-specific features for racing, shooter, and platformer games, respectively.
Genre-specific features describe events, interactions, and states from the player's perspective; i.e. only bots and objects visible to the player are logged. Specific features encode the \textit{gameplay context}, {player status}, {bot status}, and {events} both {controlled by the player} and {controlled by the game}; more details can be found at \cite{melhart2021affect}. Some games lack gameplay features that other games have, even within the same genre. In case of a missing feature, we fill the missing values with zeroes; i.e. a large loop in the track is a central feature in Solid but missing from TinyCars, subsequently features referencing the loop have a constant value of zero in TinyCars.

\begin{table}[!tb]
    \centering
    \caption{The general gameplay features of AGAIN}
    \begin{tabular}{l|l}
        {Feature} & {Description}\\
        \hline
        \hline
        Time Passed & Time since the start of the recording \\
        Player Score & Points earned by the player \\
        Input Intensity & Number of key presses \\
        Input Diversity & Number of unique key presses \\
        Player Activity & Time spent pressing controls \\
        Player Movement & Distance travelled and reticle moved \\
        Bot Count & Number of bots visible \\
        Bot Movement & Bot distance travelled \\
        Bot Diversity & Number of unique bots visible \\
        Object Intensity & Number of objects of interest \\
        Object Diversity & Number of unique objects \\
        Event Intensity & Number of events \\
        Event Diversity & Number of unique events \\
    \end{tabular}
    \label{tab:general_features}
\end{table}

Beyond these \textit{specific features}, the AGAIN dataset also provides $13$ \textit{general features} \cite{melhart2021affect}. Table \ref{tab:general_features} shows these {general features} and their short descriptions. These features describe the game on a higher level without introducing substantial domain knowledge to the data. Most general features are trivial to create, with the exception of {object intensity} and {object diversity}. What constitutes an {object} in each game varies, but in most cases, this includes passive elements the player can interact with (e.g. destructible elements and power ups).

\begin{figure}[!tb]
    \centering
    \includegraphics[width=0.94\linewidth]{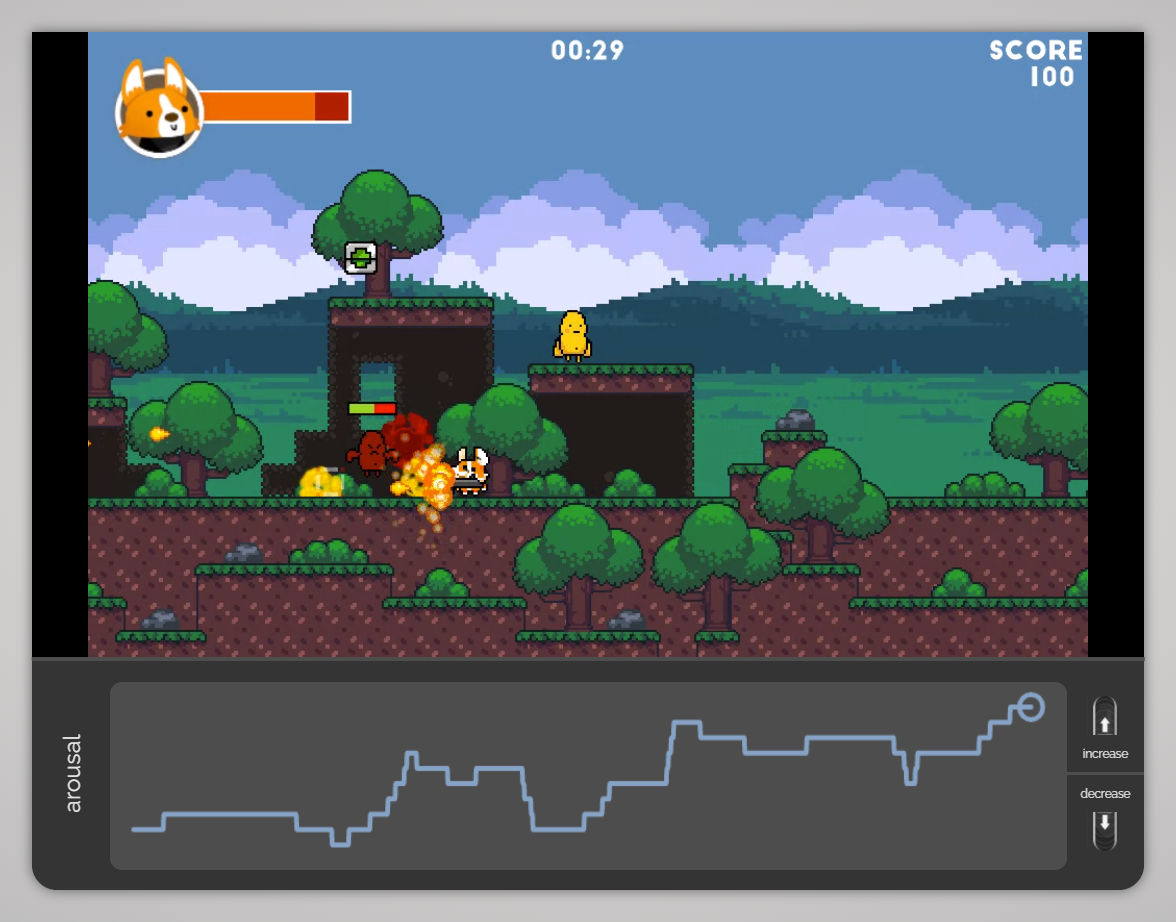}
    \caption{Continuous, unbounded arousal annotation with the RankTrace method through PAGAN \cite{melhart2021affect}. The figure shows the Run'N'Gun platformer game.}
    \label{fig:ranktrace}
\end{figure}

AGAIN offers continuous, unbounded arousal annotations of each gameplay session recorded with the PAGAN annotation tool \cite{melhart2019pagan} (see Figure~\ref{fig:ranktrace}). The interface shows the full history of the annotation process and does not limit the value range of the affect label. Due to these properties, the collected \textit{annotation trace} preserves the subjective and ordinal nature of the player experience \cite{yannakakis2018ordinal} and makes the dataset optimal for modelling through preference learning (see Section \ref{sec:methods:pl}).

\subsection{Preprocessing}

\begin{figure}[t]
    \centering
    \includegraphics[width=1\linewidth]{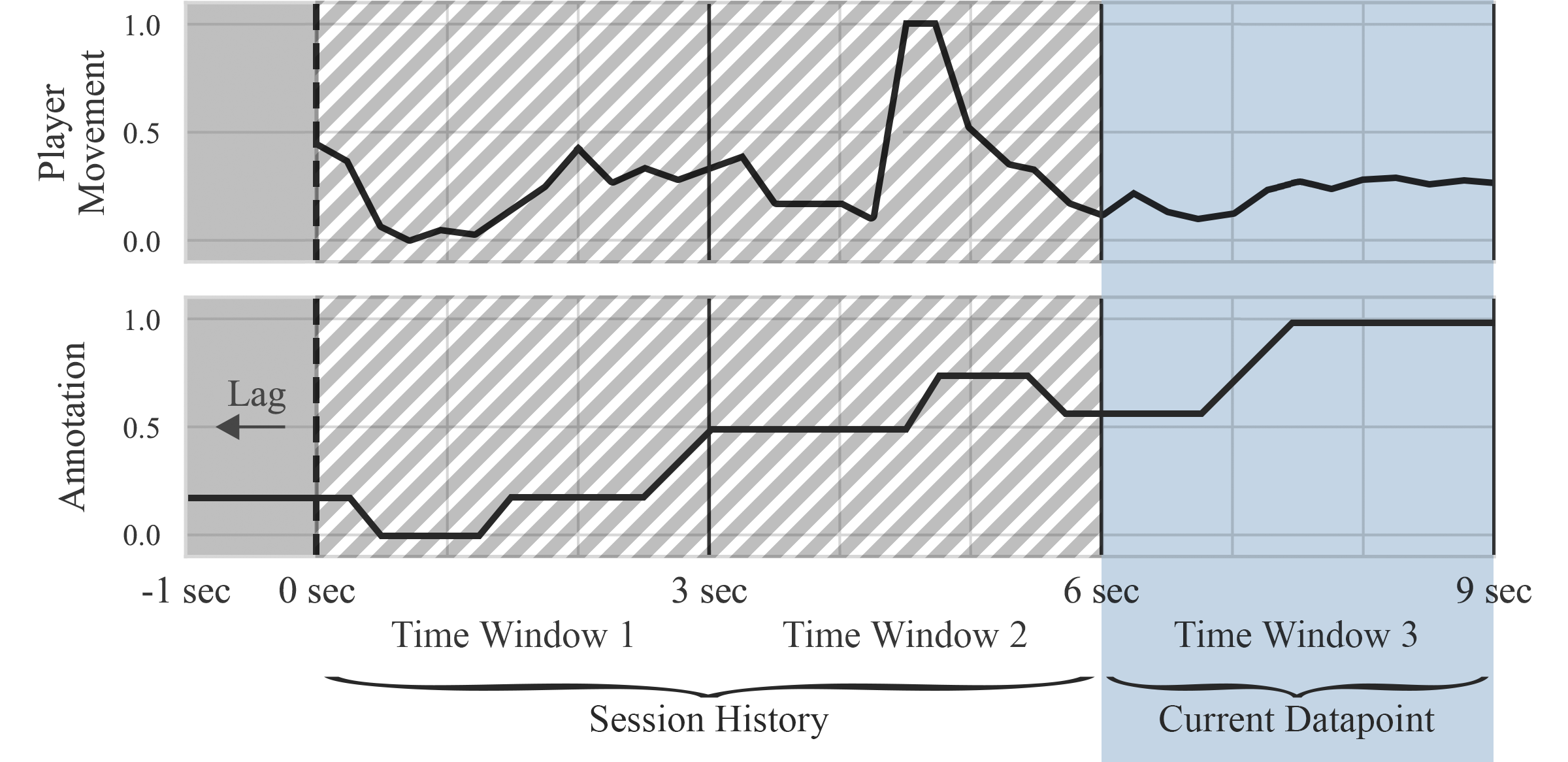}
    \caption{The aggregation of 3-second time windows. The figure shows an example of comparing the highlighted Time Window 3 to the average of the Session History (Time Window 1 and Time Window 2).}
    \label{fig:preprocessing}
\end{figure}

This paper uses the preprocessed, cleaned dataset from the AGAIN database, from which unresponsive participants and outliers have already been removed \cite{melhart2021affect}. Because the collected data is irregularly spaced due to the online collection protocol, the dataset has also been resampled at 250ms intervals \cite{melhart2019pagan,melhart2021affect}.
% This paper replicates the steps of the clean AGAIN dataset, which has been cleaned of errors, outliers, and resampled at 4Hz \cite{melhart2021affect}. 
In this section, we discuss the additional preprocessing steps we took for this paper. As windows of 250ms are not meaningful intervals in terms of human attention due to reaction time, we process the data into 3-second time windows. As presented in Section \ref{sec:background:ordinal}, we derive the mean of annotation windows. Finally, we apply a 1-second annotation lag (shifted back compared to other features) to account for reaction time. Mariooryad and Busso suggest that although an optimal annotation lag can be found algorithmically, an ad-hoc value between 1 and 3 seconds is practically a good compromise when it comes to similar annotation tasks \cite{mariooryad2013analysis}. Through a preliminary experiment, we determined that an annotation lag of 1 second is sufficient to correct for the participants' reaction. Figure \ref{fig:preprocessing} illustrates the annotation metrics and how the annotation lag is applied. After this 1-second lag correction, the dataset consists of $40,836$ datapoints.

\section{Methods}\label{sec:methods}
% \todo[color=green]{AL reviewed}

In this paper, we use preference learning (PL) to construct models of arousal. In this section we describe the core aspects of PL (Section \ref{sec:methods:pl}) and the particular algorithm used for our experiments: i.e. random forests (Section \ref{sec:methods:rf}).

\subsection{Preference Learning}\label{sec:methods:pl}

Preference learning is a supervised learning paradigm in which an algorithm learns to distinguish between datapoints in an ordinal manner \cite{furnkranz2011preference}. The name of the method stems from early applications involving recommender systems and actual user preferences \cite{joachims2002optimizing}, but PL is directly applicable to any supervised learning task in which the target outputs can be treated as ordinal data. The core of the method is the transformation of the data by discarding the output labels but conserving the relationships they describe. The algorithm then learns to predict this relationship instead of any actual label of the data. Some PL techniques derive a ranking score, which can be used for pointwise predictions. PL by \textit{pairwise comparison} has proven to be more robust, providing more stable predictions when the distribution of the labels is not normal or unknown \cite{furnkranz2003pairwise,melnikov2016pairwise}.

Formally, in {pairwise} PL for every pair of datapoints $(x, x') \in X$ and label $(\lambda_x, \lambda_{x'}) \in L$ we create two new points $(x-x')$ and $(x'-x)$ and two new labels, $y$ and $y'$. In case of $\lambda_x \succ \lambda_{x'}$ ($x$ is preferred to $x'$) we assign $y=1$ to $(x-x')$ and $y'=-1$ to $(x'-x)$, indicating the preference relation. During the $\lambda_x \succ \lambda_{x'}$ comparison a \emph{Preference Threshold} ($P_t$) parameter can be applied. $P_t$ takes a value between 0 and 1 and controls the required difference between two labels to be considered a preference. 
% The resulting dataset can grow quadratically in size depending on the valid preference relations in the dataset.
% 
While the previously published baseline of the dataset \cite{melhart2021affect} used consecutive time windows during the pairwise transformation, this paper compares datapoints to data averaged over all previous datapoints within a session. This processing method emphasises the temporality of the data by considering datapoints in relation to the session history.
The pairwise transformation is applied to each query, i.e. within each play session separately. The reformulated problem can be solved by any binary classifier. Moreover, by keeping two observations per pair, the baseline of the transformed dataset is always $50\%$. 

\subsection{Random Forests}\label{sec:methods:rf}
As explained above, through the pairwise transformation, the task of PL is reformulated as binary classification. In this paper, we use a {Random Forest} (RF) classifier as it provides a robust method for modelling arousal.
An RF is an ensemble learning method used for classification and regression. As the name suggests, RFs operate by constructing a multitude of randomly initialised independent decision trees during training and use the {mode} of their individual predictions as the meta output. Decision trees themselves are simple yet powerful machine learning algorithms for predictive modelling \cite{kratzwald2018deep}; they operate by constructing an acyclical network of nodes, which splits the features of the given dataset into simpler decisions \cite{lewis2000introduction}. For our experiments we rely on the \emph{Scikit-learn} Python library \cite{scikit-learn}. {Scikit-learn} implements decision trees through an optimised Classification And Regression Tree (CART) algorithm first proposed by Breiman \cite{breiman2017classification}. The CART method uses a generalisation of the binomial variance to evaluate the impurity (and thus splitting criterion) of nodes \cite{loh2011classification}. It also relies on a process of ``overgrowing'' and pruning trees \cite{lewis2000introduction} to minimise training errors without overfitting. We set the number of estimators to $100$ and the maximum depth of each tree to $10$ for all experiments. Because RFs are stochastic, we repeat each experiment 20 times and present the average results of all runs.

\section{Results}
% \todo[color=green]{AL reviewed, David check numbers are up to date}
% \todo[color=yellow]{D updated Sec \ref{sec:results:genre}.}

This section presents the key results of our experiments and is structured as follows. First we discuss the parameter tuning protocol (Section \ref{sec:results:tuning}) while in Section \ref{sec:results:baselines} we present the performance of \textit{game models}, i.e. models trained and validated on the same game. Finally, in Section \ref{sec:results:genre} we introduce and test \textit{genre models}, i.e. models tested on an unseen game while trained on other games of the same genre. Figure~\ref{fig:pipeline} shows an example of our pipeline when it comes to genre models.

% In all experiments reported, we present the results with the arousal annotation processed as the mean ($\mu_{A}$), which corresponds to predicting the \textit{relative change} in arousal throughout a session. 
%We measure correlation through Kendall's $\tau$, which measures monotonic relationships with greater robustness to outliers than other correlation metrics \cite{nelsen2001kendall}. 
Reported significance is measured by two-tailed Student's $t$-tests with $\alpha=0.05$, adjusted with the Bonferroni correction where applicable.

\subsection{Cross Validation and Parameter Tuning}\label{sec:results:tuning}
We use 10-fold cross-validation to test our results. The cross-validation folds are defined between subjects. Because 122 subjects cannot be divided evenly, each fold encompasses either 12 or 13 players. To make our results comparable to each other, the same cross-validation strategy is maintained with both game models and genre models. This means that in the former case, we train models on specific games and test them on unseen players of the same game, and in the latter case, we train models on two games in a given genre and test it on the \textit{unseen players of the unseen game}.

During parameter tuning, we focus on the $P_t$ parameter (see Section \ref{sec:methods:pl}), which determines which changes of arousal should be discarded as marginal. In particular, we seek the best $P_t$ parameter value between $0$ and $0.5$ with steps of $0.05$. 
% in the $P_t=\{0, 0.05, 0.1, 0.15, 0.2, 0.25\}$ space. 
Increasing $P_t$ generally leads to higher accuracies (as the separation between preferred and non-preferred classes is clearer), but there is a trade-off in the amount of discarded data. We pick the best $P_t$ given that at least 50\% of the available comparisons is maintained in the dataset. Extensive empirical experiments show that $P_t=0.15$ yields the most accurate models.

\subsection{Game Models}\label{sec:results:baselines}

% \begin{figure}[!tb]
% \centering
% \includegraphics[width=1\linewidth]{img/1games_mean.pdf}
% \caption{Baseline performance of models trained and tested on the same games. The dotted line shows the natural baseline and the error bars indicate a 95\% confidence interval. 'SotA' labels show the previously published game-specific State of the Art models on the same games.}\label{fig:baseline}
% \end{figure}

\begin{table}[!tb]
\centering
\caption{Testing accuracies (\%) of models trained and tested on the same game. Most accurate models are in bold.}
\label{tab:baseline}
\begin{tabular}{l|c|c|c}
\textbf{Game} & \textbf{Specific} & \textbf{General} & \textbf{All} \\ \hline\hline
TinyCars & \textbf{64.8$\pm$1.2} & 64.3$\pm$0.9 & 64.4$\pm$1.1 \\
Solid & 71.8$\pm$0.4 & \textbf{73.2$\pm$0.7} & 72.7$\pm$0.6 \\
ApexSpeed & 70.5$\pm$1.1 & \textbf{71.9$\pm$1.3} & 70.8$\pm$1.1 \\
\hline
Heist! & 79.4$\pm$0.6 & 79.4$\pm$0.7 & \textbf{79.8$\pm$0.7} \\
TopDown & 82.8$\pm$1.1 & 83.3$\pm$1.1 & \textbf{83.5$\pm$1.1} \\
Shootout & 85.8$\pm$0.8 & 85.8$\pm$0.8 & \textbf{85.8$\pm$0.8} \\
\hline
Endless & \textbf{69.5$\pm$1.8} & 69.1$\pm$1.8 & 68.9$\pm$1.7 \\
Pirates! & 69.5$\pm$1.6 & 68.9$\pm$1.7 & \textbf{70.0$\pm$1.7} \\
Run'N'Gun & 79.5$\pm$1.8 & 79.8$\pm$1.9 & \textbf{79.8$\pm$1.9} \\
\end{tabular}
\end{table}

Table \ref{tab:baseline} shows the test accuracies of models trained and tested on the same games.
% % \todo[inline]{David describe the table, general observations on low accuracies for 2 games, high accuracies for all shooter games}
% 
To measure the robustness of general features, we compare the genre-specific and general feature sets to each other and models using all available features. 
In $5$ out of $9$ games, a combined feature set of genre-specific and general features leads to the highest accuracies. Notably, shooter and platformer games benefit from these combined feature sets. Interestingly most racing games models based on general features outperform models based on both the combined and genre-specific feature set. While overall general features lead to better predictions than specific features, there are exceptions in TinyCars and Endless. 

It should be noted that differences between game models with different inputs are not significant and often marginal. The best performing models are trained and tested on shooter games (average accuracy of $83\%$) followed by platformers, then racing games (average accuracy of $73\%$ and $70\%$ respectively). The lack of significant difference between feature sets shows the robustness of general features in capturing the complexity of gameplay within each genre.
The lack of significant performance increase when combining the feature sets is possibly due to redundancies between genre-specific event telemetry and features such as Event Intensity and Event Diversity (see Table \ref{tab:general_features}) that accumulate gameplay events.

% Our baseline models are predicting the relative change in arousal (increase or decrease) with a high degree of accuracy (75.67\% on average across all games---see Figure \ref{fig:baseline}). Most notable is the improvement on {Shootout} compared to the previously published baseline \cite{melhart2021affect}. A closer examination of the features reveals that the \texttt{time\_passed} and \texttt{Player Score} general features yield the strongest Kendall's $\tau$ correlation with $\mu_{A}$ (0.22 and 0.26 respectively). These two features have a very high correlation ($\tau = 0.83$ for both) with the index of the given time-window within a session as well. This is unsurprising, as it has been demonstrated in \cite{melhart2021affect} that the level of arousal has a growing trend on average across all games. Because $\mu_{A}$ is highly dependent on time, features that encode time-related information improve these models greatly.

\subsection{Genre Models}\label{sec:results:genre}

After acquiring a baseline performance of game models, we move on to genre-based modelling. Each \textit{genre model} is trained on two games and tested on the remaining one within the genre; e.g. the model for TinyCars on Table \ref{tab:genre} shows the results of a model trained on {Solid} and ApexSpeed and tested on TinyCars (see Figure~\ref{fig:pipeline}). We are referring to models based on the test game in this section.

% \begin{figure}[!tb]
% \centering
% \includegraphics[width=1\linewidth]{img/2games_mean.pdf}
% \caption{Game genre-based arousal modelling: Performance of general models trained on two games and tested on an unseen game. The $x$-axis shows the game the model has been tested on. The dotted line shows the natural baseline, and the error bars indicate a $95\%$ confidence interval. 'Baseline' labels show the comparative performance of game-specific models on the same games.}\label{fig:genre}
% \end{figure}

\begin{table}[!tb]
\centering
\caption{Testing accuracies (\%) of models trained on two games and tested on an unseen game in the same genre. Most accurate models are in bold.}
\label{tab:genre}
\begin{tabular}{l|c|c|c|c}
 & \multicolumn{3}{c|}{\textbf{Genre Models}} & \textbf{Game Models} \\ \hline
\textbf{Game} & \textbf{Specific} & \textbf{General} & \textbf{All} & \textbf{Best} \\ \hline\hline
TinyCars & 66.3$\pm$1.6 & 66.2$\pm$1.5 & \textbf{66.9$\pm$1.7} & 64.8$\pm$1.2 \\
Solid & 70.6$\pm$0.6 & 72.2$\pm$0.5 & 72.3$\pm$0.6 & \textbf{73.2$\pm$0.7} \\
ApexSpeed & 67.2$\pm$1.0 & \textbf{71.9$\pm$1.4} & 69.9$\pm$1.2 & 71.9$\pm$1.3 \\
\hline
Heist! & 64.2$\pm$0.5 & 79.3$\pm$0.9 & 79.2$\pm$0.9 & \textbf{79.8$\pm$0.7} \\
TopDown & 76.3$\pm$1.0 & 83.5$\pm$1.1 & \textbf{83.7$\pm$1.1} & 83.5$\pm$1.1 \\
Shootout & 74.3$\pm$0.6 & 85.8$\pm$0.8 & 85.5$\pm$0.8 & \textbf{85.8$\pm$0.8} \\
\hline
Endless & 67.4$\pm$1.4 & \textbf{70.0$\pm$2.0} & 69.8$\pm$1.8 & 69.5$\pm$1.8 \\
Pirates! & 66.2$\pm$1.2 & 69.5$\pm$1.7 & 69.6$\pm$1.7 & \textbf{70.0$\pm$1.7} \\
Run'N'Gun & 62.1$\pm$0.9 & 74.6$\pm$1.4 & 78.0$\pm$1.7 & \textbf{79.8$\pm$1.9} \\
\end{tabular}
\end{table}

\begin{table*}[tb!]
    \centering
    \caption{Feature importance as derived from the Random Forests, averaged across games of the same genre. Features are labelled as general (G) or specific (S). Features present in top five features of all models are shown in bold.}
    \begin{tabular}{c|c|l|r||c|l|r}
        \multirow{2}{*}{\textbf{Genre}}  & \multicolumn{3}{c||}{\textbf{Game Models (averaged)}}  & \multicolumn{3}{c}{\textbf{Genre Models (averaged)}} \\
        \cline{2-7}
         & \multicolumn{2}{c|}{\textbf{Feature}}  & \textbf{Score}  & \multicolumn{2}{c|}{\textbf{Feature}}  & \textbf{Score}\\
        \hline
        \hline
        \multirow{5}{*}{\rotatebox[origin=c]{90}{Racing}}  & G & \textbf{Time Passed} & 0.089 & G  & \textbf{Time Passed}  & 0.116 \\
         & G & \textbf{Player Score} & 0.085 & G  & \textbf{Player Score}  & 0.110 \\
         & S & Player Gas Pedal & 0.062 & S  & Player Gas Pedal  & 0.066 \\
         & G & Player Activity & 0.045 & G  & Player Activity  & 0.038 \\
         & S & Bot Score & 0.033 & S  & Bot Collision  & 0.038 \\
        \hline
        \multirow{5}{*}{\rotatebox[origin=c]{90}{Shooter}}  & G & \textbf{Time Passed} & 0.167 & G  & \textbf{Time Passed}  & 0.225 \\
         & G & \textbf{Player Score} & 0.126 & G  & \textbf{Player Score}  & 0.162 \\
         & S & Bot Health & 0.054 & S  & Bot Reloading  & 0.042 \\
         & G & Bot Count & 0.051 & S  & Player Health  & 0.040 \\
         & G & Bot Diversity & 0.050 & G  & Bot Diversity  & 0.037 \\
        \hline
        \multirow{5}{*}{\rotatebox[origin=c]{90}{Platformer}}  & G & \textbf{Time Passed} & 0.106 & G  & \textbf{Time Passed}  & 0.137 \\
         & G & \textbf{Player Score} & 0.104 & G  & \textbf{Player Score}  & 0.132 \\
         & S & Player Damaged & 0.039 & S  & Player Damaged  & 0.046 \\
         & G & Bot Movement & 0.037 & S  & Player Death  & 0.035 \\
         & G & Player Movement & 0.035 & G  & Bot Movement  & 0.032 \\
    \end{tabular}
    \label{tab:feature_importance}
\end{table*}

Table \ref{tab:genre} shows the performance of our genre models.
Results reveal the robustness of general features in comparison to genre-specific ones. 
% \hl{In 4 out of the 9 games (shooter games and Run'N'Gun), models trained on genre-specific features perform significantly worse than ones trained on feature sets containing general features.} In these cases, models trained on the specific features have an average of $-11\%$ drop inaccuracy. All games where genre-specific features fail are featuring enemy projectiles and some form of shooting mechanics. Interestingly, Run'N'Gun---while it includes shooting---does not feature mouse controls, which suggests that the reason for the performance difference between genre-specific and general models is not the different control scheme but the shooter and shooter-like gameplay dynamics.
% 
In 6 out of the 9 games (all except TinyCars, Endless, and Pirates!), models trained on genre-specific features perform significantly worse than ones trained on feature sets containing general features. In these cases, models trained on the specific features have an average of $-9\%$ drop in accuracy. The effect is most prominent in games that feature enemy projectiles and some form of shooting mechanics ($-12\%$ on average). Interestingly, Run'N'Gun---while it includes shooting---does not feature mouse controls, suggesting that the reason for the performance difference between genre-specific and general models is not the different control scheme but the shooter and shooter-like gameplay dynamics. The difference in racing games is marginal ($-2\%$ on average) but still significant.

There is no significant difference between models trained on general, genre-specific and combined features. Furthermore, there is no significant difference between these models and the best game models of Section~\ref{sec:results:baselines} (included on Table \ref{tab:genre}). The average performance of the best genre models is the same as the game-specific models ($70\%$, $83\%$, and $73\%$ for the racing, shooter, and platformer games, respectively). Interestingly, models trained on Solid and ApexSpeed perform better on TinyCars than game-specific TinyCars models. While not significantly better, when predicting TinyCars, genre models show an average of $+2\%$ improvement across all feature sets, and the best fold from models trained on general features is $+11\%$ higher (up to $86\%$) than the best fold of the corresponding game model. A reason for this improvement could be that the fixed isometric view of TinyCars is interfering with the player experience, and the more conventional first- and third-person cameras of Solid and ApexSpeed provide a more consistent coupling between telemetry and arousal. 
Similarly, genre models tested on ApexSpeed, TopDown, and Endless also outperform game models trained and tested on these games, however in these cases the improvement is marginal (less than $+1\%$ on average).

\subsection{Impact of individual telemetry features}
% % \todo[inline]{David fill in}

To better understand our results and the reason behind the unexpected robustness of general features, we observe the top five most important features per genre. Feature importance is calculated as the \emph{Mean Decrease Impurity} (MDI) \cite{louppe2013understanding}, which measures the average amount by which a feature decreases the weighted impurity across all trees in the forest. The MDI value is normalised between $1$ and $0$, the latter meaning the feature is irrelevant. The ordinal importance of the features can be observed by ranking them by their corresponding MDI values. Here, we average the MDI values of features from different training folds and within a genre to get a bigger picture. Because there was no significant difference between the models trained on different feature sets, and to maximise the number of observed features, we use models trained on all (specific and general) features. 

Table~\ref{tab:feature_importance} shows the top five features in each genre ranked by their MDI values. Across all models, Time Passed and Player Score are the most important features. As Player Score is generally increasing as the game progresses, just like Time Passed, it is also a time-related feature. The prominence of these features across the board explains the robustness of genre models when compared to game models. The importance of time makes sense in the context of the games included in AGAIN as they are all designed to be casual and arcade-like. Games like these are designed with an increasing intensity. When it comes to genre models, because time-related features are game-agnostic, the more diverse datasets of two games combined possibly emphasise these features, filtering out more specific ones. The higher MDI score of Time Passed and Player Score for genre models compared to game models supports this hypothesis. 

Analysing Table \ref{tab:feature_importance} by genre, we can see that features relating to player action are more prominent in racing games. This makes sense as the competition is based more on the individual's skill than adversary play in these games. In many cases, the player swiftly overtakes the bots (or is left behind), limiting their interaction. In shooter games, both game models and genre models focus more on the bots numbers and types and the health of either their avatars or the bots. It is surprising that while for shooters there is a starker disagreement between game models and genre models in terms of feature importance, these games produced the most robust models in both cases. However, on a second inspection, this can be attributed to the exceptional prominence of time-related features within this genre. The player's status and the bot are also important in platformer games. Unsurprisingly, the health of the bots (prominent for shooting games) is replaced with the movement of the bots as anticipating the bots' position is essential for winning in this genre.

\section{Discussion}\label{sec:discussion}
% \todo[color=green]{AL reviewed}

This study presented a robust approach to general affect modelling in videogames by investigating the generality of largely game-agnostic features across three different genres. Our results show that game intensity can be modelled based on simple general features (such as score and playtime) at an accuracy comparable to models based on hand-crafted genre-specific features. These features can be used to create general models that perform comparatively to game-specific models of arousal within genres. In quick and casual games---such as those featured in the AGAIN dataset---the intensity of the gameplay increases over time to such a degree that a relatively simple algorithm can achieve up to $86\%$ average accuracy when predicting the change in player arousal. While games with shooting mechanics were easier to predict, some models leave substantial room for improvement. The least successful general models only reached up to $65\%$ when predicting the racing game TinyCars.

Unlike earlier studies \cite{shaker2015towards,camilleri2017towards}, we presented a systematic approach to the study of general affect modelling in games, investigating almost $1,000$ gameplay sessions across nine games and three genres. % For the first time, we also investigated the impact of memory window size in preference learning performance and showed an increased performance when incorporating more information into past time windows. 
% 
%We used preference learning with flexible time windows to account for user memory.
% 
Results presented in this paper show a promising path forward for general affect modelling in videogames but also highlight the challenges ahead. 
Subsequent analysis of our results showed the prominence of time-related features, which could inform future applications. Normative datasets used for research into game-playing AI use arcade-type games \cite{perez2016general}, similar to the ones included in AGAIN and used in this study. Augmenting game-playing AI with affective models could help produce more human-like agents and believable characters \cite{miles2009improving,pacheco2018studying}. Similarly, general models of affect can be used in dynamic adaptation systems, and procedural content generation in an affective loop \cite{yannakakis2018artificial}, without having to build specialised models.
The surprising robustness of the models also means that the method can possibly be extended to other domains as well, outside of game research. Time-related features can be used to build general models of any user experience with a strong temporal dynamic. Future research should focus on applications of general models of affect in other human-computer interaction applications.
While this study only focused on within-genre affect modelling, future studies should explore truly general approaches across different genres. We used top-down, hand-crafted features but the extraction of general features can be enhanced or automated by leveraging unsupervised feature extraction; transfer learning \cite{shaker2016transfer} could be useful in this direction. Since AGAIN also includes gameplay videos \cite{melhart2021affect}, the former can be achieved using deep learning, and pixel-to-affect modelling \cite{makantasis2019pixels}.

However, the study also highlights a disconnect between contemporary commercial console and computer games and arcade-type games used in different fields of game research. 
The observed trend between time and the average value of the annotation within a time window suggests a relatively easy task. As AGAIN only includes short, casual games, it is unlikely that the same results would hold for commercial games played over long periods.
The models' reliance on time-related features could mean that the presented robustness is only applicable to similarly structured experiences and would not hold up across different contemporary industrial applications. 
Future studies should aim to verify the results observed here on longer games with a shifting level of intensity. An alternative avenue for research could be on other game-agnostic features and new processing methods for the output of the models, such as the average gradient of the annotation \cite{camilleri2017towards} as it is time-independent and subsequently likely to be more robust under longer periods of play.

\section{Conclusions}\label{sec:conclusion}
% \todo[color=green]{AL reviewed}

This paper examined an approach towards general player experience modelling in a large-scale study of almost $1,000$ play sessions. Experiments focused on general models within the \emph{Racing}, \emph{Shooter}, and \emph{Platformer} genres. Results show that general features describing the player's input, the bots' actions, and the gameplay context on a high level are robust predictors of player arousal. Through two series of experiments, we created baseline game models based on genre-specific and general feature sets and genre models which pool data from two games and predict arousal of unseen players on an unseen game. Our best general models reached up to $74\%$ accuracy on average across all genres and $86\%$ at best within the shooter genre. The core findings of this paper suggest that there exist general in-game features that can predict player experience reliably and can be transferred to games of the same genre with high accuracy. The subsequent analysis of feature importance in the presented study highlights the prominence of time-related features in the machine learning models of player arousal. This result shows the importance of the temporal aspects of the player experience.

%\section*{Acknowledgments}
%This project has received funding from the EU’s Horizon 2020 programme under grant agreement No 951911, and from the University of Malta internal research grants programme Research Excellence Fund under grant agreement No 202003.

\bibliography{genre_arousal}
\bibliographystyle{IEEEtran}
\end{document}